\documentclass[12pt,twoside,bezier]{article}

\def\beq{\begin{equation}}
\def\eeq{\end{equation}}
\def\beqa{\begin{eqnarray}}
\def\eeqa{\end{eqnarray}}
\newtheorem{Th}{Theorem}
\newtheorem{Qu}{Question}
\begin{document}

\title{Statistical mechanics of interfaces}
\author{Salvador Miracle-Sol{\'e}}
\date{\normalsize\it Centre de Physique Th\'eorique, CNRS 
\break
Marseille (France)}
\maketitle

\noindent
Article for the 
{\it Encyclopedia of Mathematical Physics}, 
J.-P.\ Fran{\c c}oise, 
G.L.\ Naber and S.T.\ Tsou, 
eds.,
Elsevier, Oxford, 2006 
(ISBN 978-0-1251-2666-3), 
vol.\  5, pp.\ 55--63. 

\smallskip



\section{\bf Introduction}

When a fluid is in contact with another fluid, or with a gas,
a portion of the total free energy of the system is
proportional to the area of the surface of contact, and to a
coefficient, the surface tension, which is specific for each
pair of substances. 
Equilibrium will accordingly be obtained when the
free energy of the  surfaces in contact is a minimum. 

Suppose that we have a drop of some fluid, $b$, over 
a flat substrate, $w$, while both are exposed to air, $a$.
We have then three different surfaces of contact, and the
total free energy of the system consists of three parts,
associated to these three surfaces.
A drop of fluid $b$, will exist provided its own two surface
tensions exceed the surface tension between the substrate $w$ and the
air, i.e., provided that
$$
\tau^{{wb}}+\tau^{ba}>\tau^ {{wa}}.
$$ 
If equality is attained, 
then a film of fluid $b$ is formed, a situation which is known as
perfect, or complete wetting.

When one of the substances involved is anisotropic,   
such as a crystal, 
the contribution to the total
free energy of each element of area depends on its
orientation. 
The minimum surface free energy for a given volume determines then  
the ideal form of the crystal in equilibrium. 

It is only in recent times that equilibrium crystals 
have been produced in the laboratory,  
first, in negative crystals (vapor bubbles) of organic substances.  
Most crystals grow under non-equilibrium conditions  
and is a subsequent relaxation 
of the macroscopic crystal that restores the equilibrium.

\bigskip 


\begin{center}
\setlength{\unitlength}{12mm}

\begin{picture}(6,2.5)(0,0)
\bezier{400}(1,0)(1.5,1.5)(3,1.5)
\bezier{400}(3,1.5)(4.5,1.5)(5,0) 
\thicklines
\put(0,0){\line(1,0){6}} 
\put(4.1,1.7){$a$}
\put(2.9,.5){$b$}
\put(2.9,-.6){$w$}
\end{picture}

\medskip

\setlength{\unitlength}{12mm}
\begin{picture}(6,3)(0,0)
\put(0,.7){\line(1,0){6}} 
\thicklines
\put(0,0){\line(1,0){6}} 
\put(4.1,1.2){$a$}
\put(2.9,.3){$b$}
\put(2.9,-.6){$w$}
\end{picture}

\bigskip\bigskip\bigskip

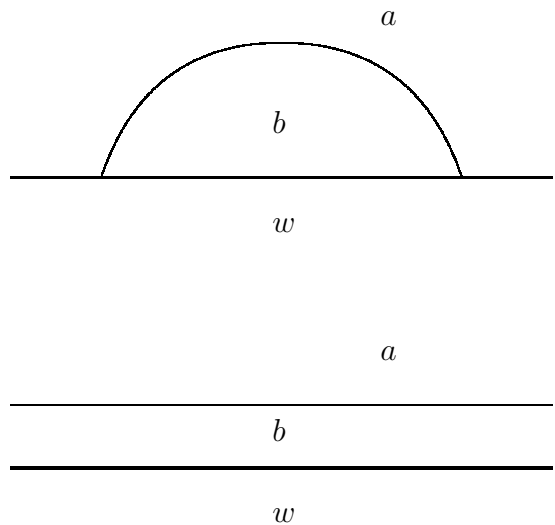
{\footnotesize Figure 1. Partial and complete wetting}

\end{center}

\bigskip


An interesting phenomenon that can be observed on 
these crystals is the roughening transition, 
characterized by the disappearance of 
the facets of a given orientation, when the temperature attains a certain 
particular value. 
The best observations have been made on helium crystals, in 
equilibrium with superfluid helium,  
since then the transport of matter and heat is extremely fast.  
Crystals grow to size of 1--5 mm and relaxation times vary 
from milliseconds to minutes. 
Roughening transitions for three different types of facets 
have been observed
(see, for instance, Wolf {\it et al.}, 1983). 

These are some classical examples among a variety of interesting 
phenomena connected with the behavior of the interface   
between two phases in a physical system. 
The study of the nature and properties of the interfaces, 
at least for some simple systems in statistical mechanics,  
is also an interesting subject of mathematical physics. 
Some aspects of this study will be discussed in the 
present article. 

We assume that the interatomic forces 
can be modelled by a lattice gas, 
and consider, as a simple example, the ferromagnetic Ising model.  
In a typical two-phase equilibrium state there is a 
dense component, which can be interpreted as a solid or liquid  
phase, and a dilute phase, which can be interpreted as the 
vapor phase. 
Considering certain particular cases of such situations,  
we first introduce a precise definition of the surface tension 
and then proceed on the mathematical analysis of 
some preliminary properties of the corresponding interfaces. 
The next topic concerns 
the wetting properties of the system,     
and the final section is devoted 
to the associated equilibrium crystal.

\section{Pure phases and surface tension}

The Ising model is defined on the 
cubic lattice ${\cal L}={\bf Z}^3$, 
with configuration space $\Omega = \{-1,1\}^{\cal L}$. 
If $\sigma\in\Omega$, the value $\sigma(i)=-1$ or $1$ is the 
spin at the site $i=(i_1,i_2,i_3)\in{\cal L}$, 
and corresponds to an empty or an occupied site in 
the lattice gas version of the model.
The system is first considered in a finite box $\Lambda\subset{\cal L}$, 
with fixed values of the spins outside. 

In order to simplify the exposition we shall mainly consider the 
three-dimensional 
Ising model, though some of the results to be discussed hold 
in any dimension $d\ge2$. 
We shall also, sometimes, refer to the two-dimensional model,   
being then understood that the definitions have been adapted 
in the obvious way.  
We assume that the box $\Lambda$ is a parallelepiped, 
centered at the origin of ${\cal L}$,
of sides $L_1,L_2,L_3$, parallel to the axes. 

A configuration of spins on $\Lambda$ 
($\sigma(i),i\in\Lambda$), 
denoted $\sigma_{\Lambda}$,  
has an energy defined by the hamiltonian 
\beq
H_{\Lambda}(\sigma_{\Lambda}\mid{\bar\sigma})
= - J \sum_{\langle i,j\rangle\cap\Lambda\not=\emptyset}
\sigma(i)\sigma(j), 
\label{A1}\eeq
where $J$ is a positive constant (ferromagnetic or attractive interaction).  
The sum runs over all nearest neighbor pairs  
$\langle i,j\rangle\subset{\cal L}$, such that at least one of the 
sites belongs to 
$\Lambda$, and one takes $\sigma(i)={\bar\sigma}(i)$  
when $i\not\in\Lambda$, the configuration ${\bar\sigma}\in\Omega$ 
being the given boundary condition. 
The probability of the configuration $\sigma_{\Lambda}$, 
at the inverse temperature 
$\beta=1/kT$, is given by the Gibbs measure 
\beq
\mu_\Lambda(\sigma_{\Lambda}\mid\bar{\sigma})=
Z^{\bar{\sigma}}(\Lambda) ^{-1} \exp
\big( -\beta H_{\Lambda}(\sigma_{\Lambda}\mid\bar{\sigma})\big), 
\label{A4}\eeq 
where $Z^{\bar\sigma}(\Lambda)$ is the partition function,  
\beq
Z^{\bar\sigma}(\Lambda)
=\sum_{\sigma_{\Lambda}}\exp \big(-\beta
H_{\Lambda}(\sigma_{\Lambda}\mid{\bar\sigma})\big). 
\label{A2}\eeq
Local properties at equilibrium can be described  
by the correlation functions between the spins on finite 
sets of sites,  
\beq
\mu_\Lambda^{\bar{\sigma}}(\sigma(A))=\sum_{\sigma_{\Lambda}}
\mu_\Lambda(\sigma_{\Lambda}\mid\bar{\sigma})
\prod_{i\in  A}\sigma(i) .
\label{A5}
\eeq 

The measures (\ref{A4}) determine (by the DLR equations) 
the set of Gibbs states of the infinite system, 
as measures on the set $\Omega$ of all configurations. 
If a Gibbs state happens to be equal to 
$\lim\mu_\Lambda(\cdot\mid\bar{\sigma})$, 
when $L_1,L_2,L_3\to\infty$, under a fixed 
boundary condition $\bar{\sigma}$, we shall call it  
the Gibbs state 
associated to the boundary condition $\bar{\sigma}$. 
One says also that this state exists in the thermodynamic limit. 
Then, equivalently, the correlation functions (\ref{A5}) 
converge to the corresponding expectation values in this state.  

This model presents, at low temperatures, that is, for
$\beta>\beta_c$, where $\beta_c$ is the critical inverse 
temperature, two different thermodynamic pure phases, 
a dense and a dilute phase in the lattice gas language 
(called here the positive and the negative phase). 
This means two extremal  
translation invariant Gibbs states, 
$\mu^+$ and $\mu^-$, 
obtained as the Gibbs states associated  
with the boundary conditions ${\bar\sigma}$, 
respectively equal to the ground configurations   
${\bar\sigma}(i) = 1$ 
and ${\bar\sigma}(i) = -1$, for all $i\in{\cal L}$. 
The spontaneous magnetization,  
\beq
m^*(\beta)=\mu^+(\sigma(i))=-\mu^-(\sigma(i)), 
\eeq 
is then strictly positive. 
On the other hand, if $\beta\le \beta_c$,  
then the Gibbs state is unique and $m^*=0$.

Each configuration inside $\Lambda$ can be described 
in a geometric way by specifying the set of Peierls contours 
which indicate the boundaries between the regions of 
spin $1$ and the regions of spin $-1$.   
Unit square faces are placed 
midway between the pairs of nearest-neighbor sites $i$ 
and $j$, perpendicularly to these bonds, whenever 
$\sigma(i)\sigma(j)=-1$. 
The connected components of this set of faces are 
the Peierls contours. 
Under the boundary conditions $(+)$ and $(-)$, 
the contours form a set of closed surfaces. 
They describe the defects of the considered configuration 
with respect to the ground states of the system 
(the constant configurations $1$ and $-1$), 
and are a basic tool for the investigation
of the model at low temperatures. 

In order to study the interface between the two pure phases
one needs to construct a state describing the coexistence
of these phases. 
This can be done by means of a new boundary condition. 
Let ${\bf n}=(n_1,n_2,n_3)$ 
be a unit vector in ${\bf R}^3$, such that $n_3>0$, and  
introduce the mixed boundary condition $(\pm,{\bf n})$, 
for which 
\beq
{\bar\sigma}(i) = 
\cases{
1  &if\quad $i\cdot{\bf n}\geq 0$, \cr
-1 &if\quad $i\cdot{\bf n}<0$. \cr}  
\label{A6}\eeq 
This boundary condition forces the system to produce a
defect going trans\-versally through the box $\Lambda$,
a big Peierls contour that can be interpreted as the  
microscopic interface (also called a domain wall).
The other defects that appear above and below the 
interface can be described by closed contours
inside the pure phases. 

The free energy, per unit area, due to the presence of the 
interface, is the surface tension. 
It is defined by
\beq
\tau({\bf n})=\lim_{L_1,L_2\to\infty}\,\lim_{L_3
\to\infty} \, -{{n_3}\over{\beta L_1L_2}}
\ln\, {Z^{\pm,{\bf n}}(\Lambda)\over Z^+(\Lambda)} 
\label{A7}\eeq 
In this expression the volume contributions 
proportional to the free energy of the coexisting phases, 
as well as the boundary effects, cancel, and only 
the contributions to the free energy due to the interface 
are left. 
The existence of such a quantity indicates that the macroscopic interface, 
separating the regions occupied by the pure phases in a large volume $\Lambda$, 
has a microscopic thickness and can therefore be 
regarded as a surface in a thermodynamic approach. 

\begin{Th} 
The interfacial free energy per unit area, $\tau({\bf n})$, 
exists, is boun\-ded, and its extension by positive homogeneity,  
$f({\bf x})=|{\bf x}|\,\tau({\bf x}/|{\bf x}|)$,  
is a convex function on ${\bf R}^3$.
Moreover, $\tau({\bf n})$ is strictly positive for
$\beta>\beta_c$, and vanishes if $\beta\le\beta_c$.

\label{T1}\end{Th}

The existence of $\tau({\bf n})$ and also the last statement 
were proved by Lebowitz and Pfister (1981), 
in the particular case ${\bf n}=(0,0,1)$,  
with the help of correlation inequalities.   
A complete proof of the theorem was given later with 
similar arguments.  
The convexity of $f$ is equivalent to the fact that
the surface tension $\tau$ satisfies a thermodynamic 
stability condition known as the pyramidal inequality 
(see Messager {\it et al.}, 1992).

\section{Gibbs states and interfaces}

In this section we consider the $(\pm,{\bf n}_0)$ boundary condition,   
also simply denoted $(\pm)$,  
associated to the vertical direction ${\bf n}_0=(0,0,1)$,     
\beq
{\bar\sigma}(i)=1\hbox{ if }i_3\ge0,\quad
{\bar\sigma}(i)=-1\hbox{ if }i_3<0.  
\label{C2}\eeq 
The corresponding surface tension is $\tau^{\pm}=\tau({\bf n}_0)$. 
We shall first recall some classical results which concern  
the Gibbs states and interfaces at low temperatures. 

According to the geometrical description of the configurations 
introduced in Section 2, we observe that 
\beq
Z^{\pm,{\bf n}}(\Lambda)/Z^{+}(\Lambda)=
\sum_{\lambda}\exp\big(-2\beta J|\lambda|-U_\Lambda(\lambda)\big),
\label{C1}\eeq 
where the sum runs over all microscopic interfaces $\lambda$ 
compatible with the boundary condition and $|\lambda|$ is 
the number of faces of $\lambda$ (inside $\Lambda$). 
The term 
$U_\Lambda(\lambda)$ equals $-\ln\, Z^+(\lambda,\Lambda)/Z^{+}(\Lambda)$,  
the sum in the partition function $Z^+(\lambda,\Lambda)$ being 
extended to all configurations whose associated contours do 
not intersect $\lambda$. 
Each term in sum (\ref{C1}) gives a weight proportional to 
the probability of the corresponding microscopic interface. 

At low (positive) temperatures, we expect the microscopic 
interface corresponding to this boundary condition,
which at zero temperature 
coincides with the plane $i_3=-1/2$,
to be modified by small deformations.
Each microscopic interface $\lambda$ can then be described   
by its defects, 
with respect to the interface at $\beta=\infty$. 
To this end one introduces some objects, 
called walls, which form the boundaries
between the horizontal plane portions of the microscopic interface, 
also called the ceilings of the interface. 

More precisely, one says that a face of $\lambda$ is a ceiling face 
if it is horizontal and such that the vertical line passing through 
its center does not have other intersections with $\lambda$. 
Otherwise, one says that it is a wall face. 
The set of wall faces splits into maximal connected components. 
The set of walls, associated to $\lambda$, is the 
set of these components, each component being identified by its 
geometric form and its projection on the plane $i_3=-1/2$. 
Every wall $\omega$, with projection $\pi(\omega)$, increases the energy 
of the interface by a quantity 
$2J\Vert\omega\Vert$, where $\Vert\omega\Vert=|\omega|-|\pi(\omega)|$,   
and two walls are compatible if their projections do not intersect.  
In this way the microscopic interfaces may be interpreted as a 
``gas of walls'' on the two-dimensional lattice. 

Dobrushin, who developed the above analysis, 
proved also the dilute character of this 
``gas'' at low temperatures. 
This implies that the microscopic interface is essentially flat, or rigid.     
One can understand this fact by noticing first, 
that the probability of a wall is less than 
$\exp(-2\beta J\Vert\omega\Vert)$, 
and second, that in order to create 
a ceiling in $\lambda$, which is not in the plane $i_3=-1/2$, one 
needs to surround it by a wall, that one has to grow when the  
ceiling is made over a larger area.  

Using correlation inequalities one proves that the Gibbs state 
$\mu^{\pm}$, associated to the $(\pm)$ boundary conditions, always 
exists, and that it is invariant under horizontal translations of the 
lattice, i.e., $\mu^{\pm}(\sigma(A+a))=\mu^{\pm}(\sigma(A))$ for 
all $a=(a_1,a_2,0)$. 
It is also an extremal Gibbs state. 
Let $m(z)$ be the magnetization $\mu^{\pm}((\sigma(z))$ at the 
site $z=(0,0,z)$.
The function $m(z)$ is monotone increasing and satisfies the 
symmetry property $m(-z)=-m(z+1)$. 
Some consequences of Dobrushin's work are the following properties. 

\begin{Th} 
If the temperature is low enough, 
i.e., if $\beta J\ge c_1$, where $c_1$ is a given constant, then  
\beqa
&& m^{\pm}(0)\hbox{ is strictly positive,}\label{C3}\\ 
&& m^{\pm}(z)\to m^*, \hbox{ when }z\to\infty,\hbox{ exponentially fast.} 
\label{C4}\eeqa 

\end{Th} 

Equation (\ref{C3}) is just another way of saying that 
the interface is rigid and that the state $\mu^{\pm}$ is  
non-translation invariant (in the vertical direction). 
Then, the correlation functions $\mu^{\pm}(\sigma(A))$ describe 
the local properties, or local structure, 
of the macroscopic interface.  
In particular, the function $m(z)$ represents the magnetization profile. 
Then statement (\ref{C4}), together with the symmetry property,  
tells us that the thickness of this interface 
is finite, with respect to the unit lattice spacing. 

The statistics of interfaces has been 
rewritten in terms of a gas of walls and  
this system may further be studied by cluster expansion 
techniques. 
There is an interaction between the walls,  
coming from the term $U_\Lambda(\lambda)$ in equation (\ref{C1}),  
but a convenient mathematical description of this interaction 
can be obtained by applying the standard low temperature 
cluster expansion, 
in terms of contours, to the regions above and below the interface. 

This method was introduced by Gallavotti in his study, mentioned below,   
of the two-dimensional Ising model.   
It has been applied by Bricmont {\it et al.},
to examine the interface structure in the present case. 
As a consequence,   
it follows that the surface tension, more exactly $\beta\tau^{\pm}(\beta)$, 
and also the correlation functions, 
are analytic functions at low temperature.  
They can be obtained as explicit 
convergent series in the variable $\zeta=e^{-2\beta J}$.  

The same analysis applied to the two-dimensional model
shows a very different behavior at low temperatures.
In this case, the microscopic interface $\lambda$ is a polygonal line 
and the walls belong to the one-dimensional lattice. 
One can then increase the size of a ceiling without modifying 
the walls attached to it.  

Indeed, Gallavotti turned this observation into a proof 
that the Gibbs state $\mu^{\pm}$ is now 
translation invariant. 
The line $\lambda$ undergoes large fluctuations 
of order $\sqrt{L_1}$, and 
disappears from any finite region of the lattice,  
in the thermodynamic limit. 
In particular, we have then  
$\mu^{\pm}=(1/2)(\mu^+ +\mu^-)$, 
a result that extends to all boundary conditions $(\pm,{\bf n})$.  

Using these results Bricmont {\it et al.} studied also the 
local structure of the interface at low temperatures and 
showed that its intrinsic thickness is finite. 
To study the global fluctuations one can compute the 
magnetization profile by introducing, 
before taking the thermodynamic limit,   
a change of scale:  
$\mu^{\pm}_\Lambda(\sigma(zL^\delta_1))$, with $\delta=1/2$ 
or near to this value.  
This is an exact computation that has been done by Abraham and Reed. 

Let us come back to the three-dimensional Ising model where  
we know that the interface 
orthogonal to a lattice axis is rigid at low temperatures.

\begin{Qu} 
At higher temperatures, but before reaching the critical temperature, 
do the fluctuations of this interface become unbounded, 
in the thermodynamic limit, 
so that the corresponding Gibbs state is translation invariant?     

\end{Qu}

One says then that the interface is rough, and  
it is believed that, effectively, the interface becomes rough  
when the temperature is raised, undergoing 
a roughening 
transition at an inverse temperature $\beta_R>\beta_c$.

It is known that $\beta_R\le\beta_c^{d=2}$, 
the critical inverse temperature of the two-dimensional Ising model, 
since van Beijeren proved using correlation inequalities, 
that above this value, the state $\mu^{\pm}$ is not 
translation invariant. 
Recalling that the rigid interface may be viewed 
as a two-dimensional system, the system of
walls, a representation that would become inappropriate 
for a rough interface, 
one might think that the phase transition  
of the two-dimensional Ising model  
is relevant for the roughening transition, and that
$\beta_R$ is somewhere near $\beta_c^{d=2}$. 
Indeed, approximate methods, used by Weeks {\it et al.}, 
give some evidence for the existence of such a $\beta_R$ 
and suggest a value slightly smaller than $\beta_c^{d=2}$, 
as shown in Table 1. 
To this day, however, it appears to be no proof of the 
fact that $\beta_R>\beta_c$, i.e., that the roughening transition 
for the three-dimensional Ising model really occurs. 

\bigskip

\begin{center}

Table 1. Some temperature values.

\bigskip

\begin{tabular}{lrl} 
\hline
$d=3$ &$\beta_c J\sim 0.22$ &approximate critical temperature \\ 
$d=3$ &$\beta_R J\sim 0.41$ &conjectured roughening temperature  \\
$d=2$ &$\beta_c J=0.44$ &exact critical temperature \\
\hline
\end{tabular}

\end{center}

\bigskip
\medskip

At present one is able to study rigorously the roughening 
transition only for some simplified models with a restricted set of 
admissible microscopic interfaces.  
Moreover, the closed contours, describing the defects above 
and below $\lambda$, are neglected, 
so that these 
two regions have the constant configurations $1$ or $-1$,    
and one has $U_\Lambda(\lambda)=0$ in equation (\ref{C1}). 

The best known of these models is the classic SOS (solid-on-solid) 
model in which the interfaces $\lambda$ have 
the property of being cut only once by 
all vertical lines of the lattice. 
This means that $\lambda$ is the graph of a function  
that can equivalently be used to define the possible 
configurations of $\lambda$. 
If $\lambda$ contains the horizontal face with center 
$(i_1,i_2,i_3-1/2)$, 
then the value at $(i_1,i_2)$ of the associated function 
is $\phi(i_1,i_2)=i_3$. 

The proof that the SOS model with the boundary condition $(\pm)$ 
has a roughening transition is a highly non-trivial result 
due to Fr\"ohlich and Spencer. 
When $\beta$ is small enough, 
the fluctuations of $\lambda$  are of order $\sqrt{\ln L}$ 
(in a cubic box of side $L$).  

Moreover, other interface models, 
with additional conditions on the allowed microscopic interfaces, 
are exactly solvable. 
The BCSOS (body-cen\-tered solid-on-solid) model, 
introduced by van Beijeren, belongs to this class.  
It is, in fact, the first model for which the existence of a 
roughening transition has been proved. 
More recently, also the TISOS (triangular Ising solid-on-solid) model, 
introduced by Bl{\"o}te and Hilhorst and further studied by 
Nienhuis {\it et al.}, 
has been considered in this context.  

The interested reader can find more information and references, 
concerning the subject of this section, in the review article 
by Abraham (1986).

\section{Wetting phenomena}

Next we consider the Ising model over  
a plane horizontal substrate (also called a wall) 
and study the difference of surface tensions 
which governs the wetting properties of this system.  

We first describe the approach developed by Fr{\"o}hlich and Pfister (1987) 
and briefly report some results of their study. 
We consider the model on the semi-infinite lattice 
\beq
{{\cal L}'}=\{i\in{\bf Z}^3 : i_3\ge0\}. 
\label{2.1}\eeq 
A magnetic field, $K\ge0$, is added on the boundary sites, 
$i_3=0$,
which describes the interaction with the substrate, 
supposed to occupy the complementary region ${\cal L}\backslash{\cal L}'$. 

We constrain the model in the finite box
$\Lambda'=\Lambda\cap{\cal L}'$, 
with $\Lambda$ as above, 
and impose the value of the spins outside. 
The hamiltonian becomes 
\beq
H_{\Lambda'}^{\rm w}(\sigma_{\Lambda'}\mid{\bar\sigma})= 
-J\sum_{\langle i,j\rangle\cap\Lambda'\ne\emptyset}\sigma(i)\sigma(j)
-K\sum_{i\in{\Lambda'},i_3=0}\sigma(i). 
\label{2.2}\eeq
Here $\sigma_{\Lambda'}$ represents the configuration inside 
$\Lambda'$, the pairs $\langle i,j\rangle$   
are contained in ${\cal L}'$,   
and $\sigma(i)={\bar\sigma}(i)$ when 
$i\not\in\Lambda'$,  
the configuration ${\bar\sigma}$ being the given boundary condition.
The corresponding partition function 
is denoted by 
$Z^{{\rm w}\bar\sigma}(\Lambda')$. 

Since there are two pure phases in the model we must consider 
two surface free energies, or surfaces tensions, 
$\tau^{{\rm w}+}$ and $\tau^{{\rm w}-}$, between the wall and 
the positive or negative phase present in the bulk. 
They are defined through the choice of the  
boundary condition, ${\bar\sigma}(i)=1$ or ${\bar\sigma}(i)=-1$, 
for all $i\in{\cal L}'$.
Let us consider first the case of the $(-)$ boundary condition. 

The surface free energy contribution, per unit area, due to the 
presence of the wall, when we have the negative phase in the bulk, is
\beq
\tau^{{\rm w}-}(\beta,K)=\lim_{L_1,L_2\to\infty}\lim_{L_3\to\infty}
-{1\over{\beta L_1L_2}}
\ln{{Z^{{\rm w}-}(\Lambda')}\over{Z^-(\Lambda)^{1\over2}}} 
\label{2.3}\eeq 
The division by $Z^{-}(\Lambda)^{1\over2}$ 
allows us to subtract from the total free energy,  
$\ln Z^{{\rm w}-}(\Lambda')$,  
the bulk term and all boundary terms which are not related to the
presence of the wall. 
The existence of limit (\ref{2.3}) follows from correlation inequalities, 
and we have $\tau^{{\rm w}-}\ge0$.

\bigskip


\setlength{\unitlength}{10mm}

\begin{center}

\begin{picture}(4,7)
\thinlines
\put(0,0){\line(1,0){3}} 
\put(0,5.5){\line(1,0){3}} 
\put(1,6.5){\line(1,0){3}} 
\put(3,0){\line(1,1){1}} 
\put(0,5.5){\line(1,1){1}} 
\put(3,5.5){\line(1,1){1}} 
\put(0,0){\line(0,1){5.5}} 
\put(3,0){\line(0,1){5.5}} 
\put(4,1){\line(0,1){5.5}} 
\multiput(1,1.1)(0,.3){18}{\line(0,1){.2}}
\multiput(.1,.1)(.1,.1){9}{.}
\multiput(1,1)(.3,0){10}{\line(1,0){.2}}
\thicklines
\put(0,2.7){\line(1,0){3}} 
\put(1,3.7){\line(1,0){3}} 
\put(0,2.7){\line(1,1){1}} 
\put(3,2.7){\line(1,1){1}} 
\put(-.5,1.5){$-$}
\put(-.5,3.9){$+$}
\put(4.2,2.5){$-$}
\put(4.2,4.9){$+$}
\end{picture}
\qquad\qquad\qquad 
\begin{picture}(4,7)
\thinlines
\put(0,0){\line(1,0){3}} 
\put(0,5.5){\line(1,0){3}} 
\put(1,6.5){\line(1,0){3}} 
\put(3,0){\line(1,1){1}} 
\put(0,5.5){\line(1,1){1}} 
\put(3,5.5){\line(1,1){1}} 
\put(0,0){\line(0,1){5.5}} 
\put(3,0){\line(0,1){5.5}} 
\put(4,1){\line(0,1){5.5}} 
\multiput(1,1.1)(0,.3){18}{\line(0,1){.2}}
\multiput(.1,.1)(.1,.1){9}{.}
\multiput(1,1)(.3,0){10}{\line(1,0){.2}}
\thicklines
\put(0,2.7){\line(1,0){1.6}} 
\put(1,3.7){\line(1,0){1.6}} 
\put(1.6,3.1){\line(1,0){1.4}} 
\put(2.6,4.1){\line(1,0){1.4}} 
\put(1.6,2.7){\line(0,1){.4}} 
\put(2.6,3.7){\line(0,1){.4}} 
\put(0,2.7){\line(1,1){1}} 
\put(3,3.1){\line(1,1){1}} 
\put(-.5,1.5){$-$}
\put(-.5,3.9){$+$}
\put(4.2,2.8){$-$}
\put(4.2,5.2){$+$}
\end{picture}

\bigskip\bigskip\bigskip\bigskip

\begin{picture}(5,4)
\thinlines
\put(0,0){\line(1,0){3}} 
\put(0,3.1){\line(1,0){3}} 
\put(1,4.1){\line(1,0){3}} 
\put(3,0){\line(1,1){1}} 
\put(0,3.1){\line(1,1){1}} 
\put(3,3.1){\line(1,1){1}} 
\put(0,0){\line(0,1){3.1}} 
\put(3,0){\line(0,1){3.1}} 
\put(4,1){\line(0,1){3.1}} 
\multiput(1,1.1)(0,.3){10}{\line(0,1){.2}}
\multiput(.1,.1)(.1,.1){9}{.}
\multiput(1,1)(.3,0){10}{\line(1,0){.2}}
\thicklines
\put(-1.5,-.5){\line(1,0){5}} 
\put(4,1.5){\line(1,0){1.5}} 
\put(-1.5,-.5){\line(1,1){1.5}} 
\put(3.5,-.5){\line(1,1){2}} 
\put(-.5,1.5){$-$}
\put(4.2,2.3){$-$}
\put(1.6,.4){$K$}
\put(4.8,.3){$W$}
\end{picture}

\end{center}

\bigskip
\bigskip 

\begin{quote}

{\footnotesize 
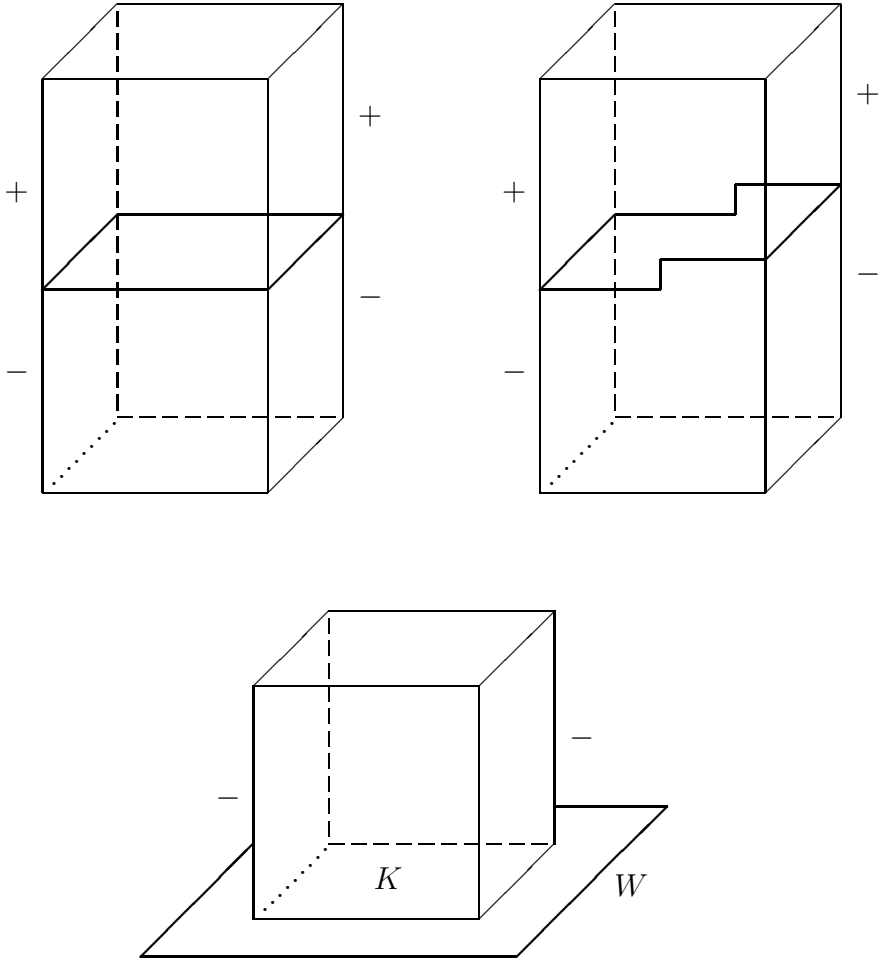
Figure 2. Boundary conditions for the cubic lattice. 
Above, the box $\Lambda$ with the $(\pm)$ and (step) boundary conditions. 
Below, the box $\Lambda'$ and the wall $W$ with the $({\rm w}-)$  
boundary conditions.}

\end{quote} 

\bigskip


One can prove, as well, the existence of the Gibbs state $\mu^{{\rm w}-}$ 
of the semi-infinite system, 
associated to the $(-)$ boundary condition. 
This state is the limit of the 
finite volume Gibbs measures $\mu_{\Lambda'}(\sigma_{\Lambda'}\mid(-))$ 
defined by hamiltonian (\ref{2.2}). 
It describes the local equilibrium properties of the system near the wall, 
when deep inside the bulk the system is in the negative phase. 
Similar definitions give the surface tension $\tau^{{\rm w}+}$ and 
the Gibbs state $\mu^{{\rm w}+}$, corresponding to the boundary condition 
$\bar\sigma(i)=1$, for all $i\in\Lambda'$.  

We remark that the states $\mu^{{\rm w}+}$ and $\mu^{{\rm w}-}$ 
are invariant by translations parallel to the plane $i_3=0$,  
and introduce the magnetizations,  
$m^{{\rm w}-}(z)=\mu^{{\rm w}-}(\sigma(z))$, 
where $z$ denotes the site $(0,0,z)$, 
$m^{{\rm w}-}=m^{{\rm w}-}(0)$, and similarly $m^{{\rm w}+}(z)$ and 
$m^{{\rm w}+}$. 
Their connection with the surface free energies is given by the formula 
\beq
\tau^{{\rm w}-}(\beta,K)-\tau^{{\rm w}+}(\beta,K)=
\int_0^K (m^{{\rm w}+}(\beta,s)-m^{{\rm w}-}(\beta,s))ds.
\label{f}\eeq

We mention in the following theorem some results of 
Fr{\"o}hlich and Pfister's study. 
Here $\tau^{\pm}$ is, as before, the usual surface tension  
between the two pure phases of the system, for a  
horizontal interface. 

\begin{Th}
With the above definitions, we have
\beqa
\tau^{{\rm w}-}(\beta,K)-\tau^{{\rm w}+}(\beta,K)
&\le&\tau^{\pm}(\beta), \label{t}\\ 
m^{{\rm w}+}(\beta,K)-m^{{\rm w}-}(\beta,K)&\ge&0 \label{m},
\eeqa 
and the difference in (\ref{m}) is a monotone decreasing function of 
the parameter $K$.  
Moreover, if $m^{{\rm w}+}=m^{{\rm w}-}$, then the Gibbs states 
$\mu^{{\rm w}+}$ and $\mu^{{\rm w}-}$ coincide. 

\label{T3}\end{Th}

The proof is a subtle application of correlation inequalities. 
Since, from Theorem \ref{T3},   
the integrand in equation (\ref{f}) is a positive and decreasing function, 
the difference $\Delta\tau=\tau^{{\rm w}-}-\tau^{{\rm w}+}$ 
is a monotone increasing and concave (and hence continuous) 
function of the parameter $K$. 
On the other hand, one can prove that
$\Delta\tau=\tau^{\pm}$, if $K\ge J$.  
This justifies the following definition 
\beq
K_w(\beta)=\min\{\,K:\Delta\tau(\beta,K)=\tau^{+-}(\beta)\}.
\label{19}\eeq

In the thermodynamic description of wetting, the partial wetting regime 
is characterized by the strict inequality in equation (\ref{t}). 
Equivalently, by $K<K_w(\beta)$. 
We must have then $m^{{\rm w}+}\ne m^{{\rm w}-}$, 
because of equation (\ref{f}). 
This shows that, in the case of partial wetting,  
$\mu^{{\rm w}+}$ and $\mu^{{\rm w}-}$ are different Gibbs states. 

The complete wetting regime is characterized by the equality in equation 
(\ref{t}), that is, by $K\ge K_w(\beta)$. 
Then, we have $m^{{\rm w}+}=m^{{\rm w}-}$, and taking into account 
the last statement in Theorem \ref{T3}, 
also $\mu^{{\rm w}+}=\mu^{{\rm w}-}$. 
This last result implies that there is only one Gibbs state. 
Thus complete wetting corresponds to unicity of the Gibbs state. 

In this case, we also have $\lim\, m^{{\rm w}-}(z)=m^*$,  
when $z\to\infty$, because this is always true for $m^{{\rm w}+}(z)$.   
This indicates that we are in the positive phase of the system although 
we have used the $(-)$ boundary condition, so that the bulk negative phase 
cannot reach anymore the wall. 
The film of positive phase, which wets the wall completely, 
has an infinite thickness with respect to the unit lattice spacing, 
in the thermodynamic limit. 

When $\beta=\infty$ only few particular ground configurations
contribute to the partition functions, such as the configuration
$\sigma(i)=-1$ for the partition function $Z^{{\rm w}-}$, etc.,
and we obtain $\Delta\tau=2K$ and $\tau^{\pm}=2J$.
For non zero but low temperatures the small perturbations
of these ground states have to be considered,
a problem that can be treated by the method of cluster
expansions.
In fact, the corresponding defects can be described by closed 
contours as in the case of pure phases. 

\begin{Th} 
For $K<J$, the functions 
$\beta\tau^{{\rm w}-}(\beta,K)$ and $\beta\tau^{{\rm w}+}(\beta,K)$ are  
analytic at low temperatures, i.e., 
provided that $\beta\,(J-K)\ge c_2$, where $c_2$ is a given constant. 
Moreover, $m^{{\rm w}+}(z)$ and $m^{{\rm w}-}(z)$ tend, respectively,  
to $m^*$ and to $-m^*$, when $z\to\infty$, exponentially fast.

\label{T4}\end{Th} 

The last statement in Theorem \ref{T4} tells us that the wall affects only 
a layer of finite thickness 
(with respect to the lattice spacing). From a macroscopic point of view
the negative phase reaches the wall, and we are in the partial wetting regime. 
Indeed, a strict inequality holds in equation (\ref{t}).
 
Thus, for $K<J$ there is always partial wetting at low temperatures.
Then the following question arises: 

\begin{Qu}  
Is there a situation of complete wetting at higher temperatures ? 
It is understood here that $K$ takes a fixed value,  
characteristic of the substrate, such that $0<K<J$. 

\end{Qu} 

This is known to be the case in dimension $d=2$, where the 
exact value of $K_{w}(\beta)$ can be obtained from Abraham's  
solution of the model:  
$$
\cosh 2\beta K_w=\cosh 2\beta J-e^{-2\beta J}\sinh 2\beta J. 
$$
Then complete wetting occurs for $\beta$  
in the interval $\beta_c<\beta\le\beta_{w}(K)$, 
where $\beta_c$ is the critical inverse temperature 
and $\beta_{w}(K)$ is the solution of $K_{w}(\beta)=K$. 
The case $d=2$ has been reviewed in Abraham (1986). 

To our knowledge, the above question remains 
an open problem for the Ising model in dimension $d=3$. 
The problem has however been solved for the simpler case of a 
SOS interface model. 
In this case, 
a nice and rather brief proof has been given by Chalker (1982) 
of the following result:  
One has $m^{{\rm w}+}=m^{{\rm w}-}$, and hence complete wetting, if 
$$
2\beta(J-K)<-\ln(1-e^{-8\beta J}).
$$ 

It is very plausible that a similar statement is valid 
for the semi-infinite 
Ising model and, also, that Chalker's method could play a role for 
extending the proof to this case, provided an additional 
assumption is made. 
Namely, that $\beta$ is sufficiently large, and hence $J-K$ 
small enough, in order to insure the convergence of the cluster 
expansions and to be able to use them.

\section{Equilibrium crystals} 

The shape of an equilibrium crystal is obtained,
according to thermodynamics,
by minimizing the surface 
free energy between the crystal and the medium, 
for a fixed volume of the crystal phase. 
Given the orientation dependent surface tension $\tau({\bf n})$, 
the solution to this variational problem, known under
the name of Wulff construction, is the following set:
\beq
{\cal W}=\{{\bf x}\in{\bf R}^3 : 
{\bf x}\cdot{\bf n}\le\tau({\bf n})\hbox{ for all } 
{\bf n}\}.
\label{E1}\eeq
Notice that the problem is scale invariant, so that if we solve
it for a given volume of the crystal, we get the 
solution for other volumes by an appropriate scaling.
We notice also that 
the symmetry $\tau({\bf n})=\tau(-{\bf n})$
is not required for the validity of formula (\ref{E1}).  
In the present case $\tau({\bf n})$ 
is obviously a symmetric function, but 
non-symmetric situations are also physically interesting and appear, 
for instance, in the case of 
a drop on a wall discussed in Section 4. 

The surface tension in the Ising model, 
between the positive and negative phases, 
has been defined in equation (\ref{A7}). 
In the two-dimensional case, 
this function $\tau({\bf n})$ 
has (as shown by Abraham) an exact expression in terms of 
some Onsager's function. 
It follows (as explained in Miracle-Sole, 1999) 
that the Wulff shape ${\cal W}$,  
in the plane $(x_1,x_2)$, is give by 
$$
\cosh\beta x_1+\cosh\beta x_2\le\cosh^2 2\beta J/\sinh2\beta J.  
$$ 
This shape reduces to the empty set for $\beta\le\beta_c$, 
since the critical $\beta_c$ satisfies $\sinh2J\beta_c=1$. 
For $\beta>\beta_c$, it is a strictly convex set with smooth 
boundary. 

\bigskip 
\bigskip


\begin{center}

\setlength{\unitlength}{1mm}
\begin{picture}(40,60)
\thinlines
\put(0,15.3){\line(4,-3){18.5}}
\put(20,55){\line(4,-3){19.25}}
\put(21.5,1.3){\line(4,3){18.5}}
\put(.75,40.5){\line(4,3){19.25}}
\put(0,15.3){\line(0,1){22.7}}
\put(40,15.2){\line(0,1){22.8}}
\put(20,55){\line(0,1){7}}
\put(40,15.2){\line(4,-3){5}}
\put(0,15.3){\line(-4,-3){5}}
\thicklines
\bezier{200}(.75,40.5)(16.25,29.5)(20,29.5)
\bezier{200}(0,38)(15,26)(16.5,23)
\bezier{200}(16.5,23)(18,20)(18.25,1.5)
\bezier{200}(20,29.5)(23.75,29.5)(38.75,40.5)
\bezier{200}(23.5,23)(25,26)(39.75,38)
\bezier{200}(21.5,1.5)(22,20)(23.5,23)
\bezier{30}(18.5,1.5)(20,.5)(21.5,1.5)
\bezier{30}(0,38)(.2,40)(.75,40.5)
\bezier{30}(39,40.5)(39.5,40)(39.75,38)
\put(19,37){$3$}\put(9,17){$1$}\put(29,17){$2$}
\end{picture}

\end{center}

\begin{quote}

{\footnotesize 
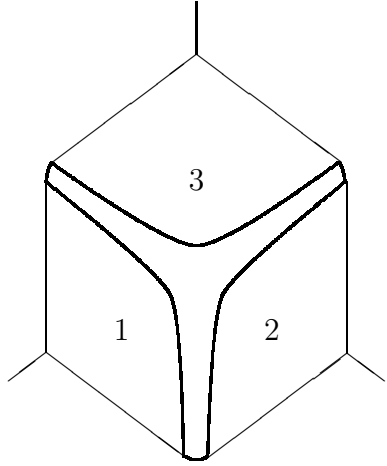
Figure 3. Cubic equilibrium crystal 
shown in a projection parallel to the (1,1,1) direction.
The three regions 1), 2) and 3) indicate the facets  
and the remaining area represents a curved part 
of the crystal surface.}

\end{quote}

\bigskip


In the three-dimensional case, 
only certain interface models can be exactly solved 
(see Section 3). 
Consider the Ising model at zero temperature. 
The ground configurations have only one defect, 
the microscopic interface $\lambda$, 
imposed by the boundary condition $(\pm,{\bf n})$. 
Then, from equation (\ref{C1}), we may write 
\beq
\tau({\bf n})=\lim_{L_1,L_2\to\infty}{{n_3}\over{L_1L_2}}
\,\big(E_\Lambda({\bf n})-\beta^{-1}N_\Lambda({\bf n})\big),
\label{DD}
\eeq 
where $E_\Lambda=2J|\lambda|$ is the energy (all $\lambda$ 
have the same minimal area) and $N_\Lambda$ the number of 
the ground states. 
Every such $\lambda$ has the property of being cut only once by 
all straight lines orthogonal to the diagonal plane 
$i_1+i_2+i_3=0$,   
provided that $n_k>0$, for $k=1,2,3$.  
Each $\lambda$ can then be described by an integer function defined on 
a triangular plane lattice, 
the projection of the cubic lattice ${\cal L}$ 
on the diagonal plane. 
The model defined by this set of admissible microscopic interfaces 
is precisely the TISOS model.  
A simi\-lar definition can be given for the BCSOS model that 
describes the ground configurations 
on the body-centered cubic lattice. 

{}From a macroscopic point of view,
the roughness or the rigidity of an interface should be apparent
when considering the shape of the equilibrium crystal 
associated with the system.
A typical equilibrium crystal at low 
temperatures has smooth plane facets 
linked by rounded edges and corners. 
The area of a particular facet decreases as the 
temperature is raised and the facet finally disappears 
at a temperature characteristic of its orientation.
It can be argued  
that the disappearance of the facet corresponds to 
the roughening transition of the interface
whose orientation is the same 
as that of the considered facet.

The exactly solvable interface models mentioned above,
for which the function $\tau({\bf n})$ 
has been computed,
are interesting examples of this behavior, 
and provide a valuable information on several aspects 
of the roughening transition. 
This subject has been reviewed by 
Abraham (1986), van Beijeren and Nolden (1987),   
and Kotecky (1989). 

For example, we show in figure 2 the shape predicted by the TISOS 
model (one eighth of the shape because of the condition $n_k>0$). 
In this model the interfaces orthogonal to the three coordinate axes 
are rigid  at low temperatures.  

For the three-dimensional Ising model at positive temperatures, 
the description of the microscopic interface, for any 
orientation ${\bf n}$, 
appears as a very difficult problem.   
It has been possible, however, to analyze 
the interfaces which are very near 
to the particular orientations ${\bf n}_0$,  
discussed in Section 3.
This analysis allows us to determine
the shape of the facets in a rigorous way. 

We first observe that 
the appearance of a facet 
in the equilibrium crystal shape is related, 
according to the Wulff construction, to the existence 
of a discontinuity in the derivative of the surface
tension with respect to the orientation. 
More precisely, 
assume that the surface tension satisfies
the convexity condition of Theorem \ref{T1}, and let 
this function $\tau({\bf n})=\tau(\theta,\phi)$
be expressed in terms of the spherical coordinates
of ${\bf n}$, 
the vector ${\bf n}_0$ being taken as the $x_3$ axis.
A facet orthogonal to ${\bf n}_0$ appears
in the Wulff shape 
if, and only if, the derivative
$\partial\tau(\theta,\phi)/\partial\theta$
is discontinuous at the point $\theta=0$,
for all $\phi$. 
The facet ${\cal F}\subset\partial{\cal W}$ 
consists of the points ${\bf x}\in{\bf R}^3$
belonging to the plane $x_3=\tau({\bf n}_0)$ and such that, 
for all $\phi$ between $0$ and $2\pi$,  
\beq
x_1\cos\phi+x_2\sin\phi\le
\partial\tau(\theta,\phi)/\partial\theta\,\vert\,_{\theta=0^+}  
\label{B1}\eeq

The step free energy is expected to play an important role 
in the facet formation.
It is defined as the free energy 
associated with the introduction of a step of height 1   
on the interface,  
and can be regarded as an order parameter
for the roughening transition. 
Let $\Lambda$ be a parallelepiped as in Section 2, 
and introduce the 
$({\rm step},{\bf m})$
boundary conditions, associated to the unit vectors
${\bf m} = (\cos\phi,\sin\phi)\in{\bf R}^2$, by
\beq
{\bar\sigma}(i) = 
\cases{
1  &if\ \ $i>0$\ \ or if\ \ $i_3=0$\ \ and\ \ $i_1m_1+i_2m_2\ge0$,\cr
-1  &otherwise. \cr}  
\label{B2}\eeq
Then, the step free energy, per unit length, for a step
orthogonal to ${\bf m}$ (with $m_2>0$) 
on the horizontal interface, is   
\beq 
\tau^{\rm step}(\phi) =
\lim_{L_1\to\infty}\lim_{L_2\to\infty}\lim_{L_3\to\infty}
- {{\cos\phi}\over{\beta L_1}}\  
\ln\  {{Z^{{\rm step},{\bf m}}(\Lambda)}\over 
{Z^{\pm,{\bf n}_0}(\Lambda)}} 
\label{B3}\eeq

A first result concerning this point,   
was obtained by Bricmont {\it et al.}, 
by proving 
a correlation inequality which establish
$\tau^{\rm step}(0)$ 
as a lower bound to the one-sided derivative
$\partial\tau(\theta,0)/\partial\theta$ at $\theta=0^+$
(the inequality extends also to $\phi\ne0$).     
Thus when $\tau^{\rm step}>0$ a facet is expected. 

Using the perturbation theory of the horizontal interface,  
it is possible to study also the 
microscopic interfaces associated with the 
$(\hbox{\rm step},{\bf m})$ boundary conditions. 
When considering these configurations, 
the step may be viewed as an additional defect
on the rigid interface described in Section 2. 
It is, in fact,
a long wall going from one side to the other side 
of the box $\Lambda$. 
The step structure at low temperatures can then be 
analyzed with the help of a new cluster expansion. 
As a consequence of this analysis we have 
the following theorem.

\begin{Th} 
If the temperature is low enough,  
i.e., if $\beta J\ge c_3$, where $c_3$ is a given constant, 
then the step free energy, $\tau^{\rm step}(\phi)$,
exists, is strictly positive, and 
extends by positive homogeneity to a strictly convex function.
Moreover, $\beta\tau^{\rm step}(\phi)$ is an
analytic function of $\zeta=e^{-2J\beta}$, 
for which an explicit convergent series expansion can be found.

\label{TT1}\end{Th}

Using the above results on the step structure, 
similar methods allow us to evaluate
the increment in surface tension of an interface 
titled by a very small angle $\theta$ 
with respect to the rigid horizontal interface. 
This increment can be expressed in terms of the 
step free energy and one obtains the following relation. 
                                           
\begin{Th} 
For $\beta J\ge c_3$, we have 
\beq
\partial\tau(\theta,\phi)/\partial\theta\,\vert\,_{\theta=0^+}
= \tau^{\rm step}(\phi).   
\label{B4}\eeq

\label{TT2}\end{Th}

This relation, together with equation (\ref{B1}),
implies that one obtains the shape of the facet 
by means of the two-dimensional Wulff construction
applied to the step free energy.
The reader will find a detailed discussion on these points, 
as well as the proofs of Theorems \ref{TT1} 
and \ref{TT2}, in Miracle-Sole (1995). 

{}From the properties of $\tau^{\rm step}$ stated in Theorem \ref{TT1}, 
it follows that the Wulff equilibrium crystal 
presents well defined boundary lines, smooth and  
without straight segments, between
a rounded part of the crystal surface and the facets parallel
to the three main lattice planes.

It is expected, but not proved, that at a higher temperature, 
but before reaching the critical temperature, 
the facets associated with the Ising model undergo a 
roughening transition. 
It is then natural to believe that the equality (\ref{B4}) 
is true for any $\beta$ larger than $\beta_R$,   
allowing us to determine the facet shape from  
equations (\ref{B1}) and (\ref{B4}), 
and that for $\beta\le\beta_R$,  
both sides in this equality vanish,
and thus, the disappearance of the facet is involved.
However, the condition that the temperature 
is low enough is needed in the proofs of Theorems \ref{TT1} 
and \ref{TT2}. 


\medskip
\medskip

\noindent
{\large\sl See also:}
{\sf Gibbs states. Ising model. Phase transitions. Wulff droplets.} 

\section{Further reading}

\parskip 8pt
\parindent 0pt
\small

Abraham, D.B. (1986) 
Surface structures and phase transitions. 
In: Domb, C. and Lebowitz, J.L. (eds.) 
{\it Critical Phenomena}, 
vol. 10, pp 1--74.
Academic Press, London.

Chalker, J.T. (1982) 
The pinning of an interface by a planar defect.
{\it Journal of Physics A: Mathematical and General}  
15, L481--L485. 

Fr{\"o}hlich, J. and Pfister, C.E. (1987) 
Semi-infinite Ising model: I and II. 
{\it Communications in Mathematical Physics} 
109, 493--523 and 
112, 51--74.

Gallavotti, G. (1999) 
{\it Statistical mechanics: a short treatise}. 
Springer, Berlin. 

Kotecky, R. (1989) 
Statistical mechanics of interfaces and equilibrium 
crystal shapes. 
In: Simon, B., Truman, A., and Davies, I.M. (eds.)  
{\it IX International Congress of Mathematical Physics},
pp 148--163,
Adam Hilger, Bristol.

Lebowitz, J.L. and Pfister, C.E. (1981) 
Surface tension and phase coexistence.
{\it Physical Review Letters} 
46, 1031--1033. 

Messager, A., Miracle-Sole, S., and Ruiz, J. (1992) 
Convexity properties of the surface tension and 
equilibrium crystals.
{\it Journal of Statistical Physics}
67, 449--470.  

Miracle-Sole, S. (1995)  
Surface tension, step free energy and facets in the
equilibrium crystal shape. 
{\it Journal of Statistical Physics}
79, 183--214. 

Miracle-Sole, S. (1999)  
Facet shapes in a Wulff crystal.   
In: Miracle-Sole, S., Ruiz, J., Zagrebnov,  V. (eds.)  
{\it Mathematical Results in Statistical Mechanics},  
pp 83--101. 
World Scientific, Singapore.  

van Beijeren, H. and Nolden, I. (1987)
The roughening transition. 
In: Schommers, W. and von Blackenhagen, P. (eds.) 
{\it Topics in Current Physics},  
vol. 43, pp 259--300. 
Springer, Berlin.

Wolf, P.E., Balibar, S., and Gallet, F. (1983) 
Experimental observations of a third roughening 
transition in hcp $^4$He crystals. 
{\it Physical Review Letters} 
51, 1366--1369.

\end{document}